# Size-enhanced, coherent photoluminescence from CdS-ZnO nanocomposite thin film

Pushan Ayyub,[1] Parinda Vasa,[1] Praveen Taneja,[1] Rajarshi Banerjee[2] and Bhanu P. Singh[1]

[1]Tata Institute of Fundamental Research, Mumbai, India

[2]Ohio State University, Columbus, Ohio

[3]Indian Institute of Technology Mumbai, India

**Abstract**

We show that the photoluminescence emitted from a dense, two-component ensemble of quantum dots is significantly higher than that from quantum dots of either of the two pure systems (CdS and ZnO). The semiconductor nanocomposite, in which the characteristic grain size of each species was 2-3 nm, was deposited directly on Si wafers by high-pressure magnetron sputtering. It exhibits a single, relatively sharp optical absorption edge. Further, using the classical Young's double slit experiment we show – possibly for the first time - that the emitted photoluminescence has significant spatio-temporal coherence, a fact that would be crucial for their use in quantum information processing as well as for lasing action from light-emitting particles.

Corresponding Author: Pushan Ayyub (pushan@mailhost.tifr.res.in)

# 1. Introduction

Though the electronic and optical properties of quasi-zero dimensional semiconductors or quantum dots is fairly well understood, there is still an overwhelming interest in these systems, largely due to the size-controlled tunability of their band-gap [1] and the high quantum efficiency for photoluminescence (PL) [2] arising from quantum confinement. Exciting applications are envisaged for such systems in optoelectronic devices, lasers, optical communications and quantum computing. Of particular interest is the lasing action arising from photon localization in an ensemble of light-emitting particles. Repeated scattering of light from grain boundaries in such systems could result in random lasing [3]. Since lasing originates from interference effects, it is essential for the emitted light to have some degree of coherence. Coherence is also important for applications related to quantum-information processing [4]. Surprisingly, there appear to be no quantitative measurements of the degree of coherence in the PL output from nanoparticles. In this letter we show that the PL emitted by an optically flat, quantum dot solid (QDS) thin film deposited on a Si wafer shows a significant enhancement due to quantum confinement. Further, using the celebrated Young's double-slit experiment [5], we show that the light emitted from such a nanocrystalline ensemble has an appreciable degree of spatial coherence. It is important to note that while the optoelectronic properties of low-dimensional semiconductors have been investigated widely [6,7] this is one of the few studies on dense ensembles of quantum dots or quantum dot solids [8,9].

A major problem with isolated semiconductor nanoparticles is that they often have surface electronic states within the HOMO-LUMO gap that provide non-radiative decay channels and lead to a severe degradation in optoelectronic properties. This problem has been addressed by 'capping' the semiconductor nanoparticle with a thin layer of a higher band gap material [10]. Techniques involving semiconductor-doped glasses [11], zeolite encapsulation [12], sol-gel synthesis [13] and colloidal precipitation [14] can produce capped, nearly monodisperse nanoparticles suitable for basic studies, but are not very useful for fabricating large area thin films that can be integrated into devices. Here, we demonstrate a *single-step* process (without the necessity of forming a core-shell structure) for the deposition of nanocomposite thin films directly on Si wafers, and show that the larger band-gap component of the nanocomposite effectively passivates the component with the smaller band-gap.

## 2. Synthesis.

Bulk CdS is a direct band gap semiconductor ($E_g$ = 2.42eV at 300K) with a hexagonal wurtzite structure. It undergoes a size-induced structural transformation to a cubic zinc blende structure below 2–3nm [15]. The larger band gap component in our nanocomposite was ZnO ($E_g$ = 3.44eV at 300K), a semiconductor with a crystal structure similar to CdS. Nanostructured films (average grain size ≈2–20nm) of most metals, semiconductors and oxides can be sputtered onto desired substrates by dc/rf sputtering [16] at relatively high gas pressures (≈10–200 mTorr) and low substrate temperatures (77–300K). In particular, uniform, nanocrystalline films of pure CdS can be synthesized by this technique [17], but need to be chemically passivated by a prolonged *in-situ* exposure to $H_2+N_2$ gas mixture. The present study reports the deposition of CdS+ZnO nanocomposite thin films on Si wafers or quartz plates by rf- magnetron sputtering from a sintered target composed of a mixture of CdS and ZnO. The sputtering was typically carried out in flowing Ar (99.999%) at 170 mTorr, the substrate being held at 0°C.

## 3. Microstructure

The x-ray diffraction (XRD) spectrum of the sputter-deposited nanocomposite film shows a single broad hump, which could arise from a superposition of the stronger diffraction lines of CdS and ZnO, broadened due to small particle size. The microstructure and chemical composition of the ion-milled nanocomposite CdS+ZnO film were obtained using transmission electron microscopy (TEM) in the imaging and diffraction modes, respectively. Bright-field TEM micrographs of the nanocrystalline CdS+ZnO film (Fig. 1, top) indicate that the average in-plain grain size is $\approx$ 2-3 nm and the size distribution is quite narrow. The almost mono-dispersed, individual nanocrystals can be more clearly identified in dark field TEM micrographs (Fig. 1, center) formed by intensity contributions from a single diffraction ring. It is not possible to obtain images at higher resolution since this results in electron beam induced annealing of the densely packed nanoparticles. *All* the distinct rings in the selected area diffraction (SAD) pattern from the sample (Fig 1, bottom) can be indexed consistently to the hexagonal and cubic polymorphs of CdS and ZnO. Energy dispersive x-ray spectrometry indicated that the CdS and ZnO phases were approximately stoichiometric, with Cd:S = 1:1.20 and Zn:O = 1:1.18.

The nanocomposite CdS+ZnO thin films therefore consist of densely packed, randomly intermixed, nearly uniform-sized arrays of *individual* CdS and ZnO nanoparticles, the characteristic grain size for each phase being 2-3 nm. There is no evidence for a core-shell structure (neither is such a structure expected in a sputtering process); or for the formation of a solid solution between CdS and ZnO. Since the ZnO phase (whose HOMO-LUMO gap is quantum shifted deep into the UV) essentially provides a passivating matrix (see below), the system can be modeled as a random, 3D ensemble of nearly identical CdS quantum dots. Such a system has been termed a disordered quantum dot solid [8] and is described by coexisting discrete (localized) and band-like (delocalized) electronic states.

**4. Optical absorption.**

The optical absorption spectrum of a typical nanocomposite CdS+ZnO thin film deposited on quartz shows a single, sharp absorption edge at 380 nm (Fig. 2, inset), which corresponds to the quantum shifted excitonic absorption energy in nano-CdS. Using standard expressions [18], we find that the observed blue shift in the gap energy (from 514nm to 380nm) corresponds to a particle diameter of 2.8nm for CdS nanoparticles, which is in close accordance with the size obtained from TEM. Clearly, the observed absorption edge is not related to HOMO-LUMO gap in ZnO nanoparticles, which is expected at ≈320nm for ≈3nm particles, its "bulk" band gap being 360nm.

## 5. Photoluminescence (PL)

The PL emission spectra were obtained using the 457.9 nm excitation of an Argon laser, a 0.85m double monochromator (with excellent stray light rejection), and a cooled photomultiplier. Figs. 2(a) and 2(b) show the PL emission spectra from CdS+ZnO nanocomposite thin films of different thickness but similar particle size, while Figs. 2(c) and 2(d) show those from pure CdS nanocrystalline thin films. The thinner CdS film (Fig. 2d) consists of smaller particles (2.4nm) and shows a broad band PL, while the thicker CdS film (Fig. 2c) consists of larger particles (5.2nm) and shows a much weaker PL due to reduced confinement effects. Note that the peak of the PL emission profile occurs at ≈560nm in nanocrystalline CdS as well as CdS-ZnO nanocomposite, indicating a similar physical origin. The maximum in the PL emission is 'Stokes-shifted' with respect to the absorption edge in CdS and CdS+ZnO due to trapping and capture in defect states, and is a favourable feature in many applications [19]. Fig 2(e) shows the PL from a nanocrystalline ZnO thin film with a mean particle size of 3nm. The visible (green) emission in bulk as well as nanocrystalline ZnO involves recombination between oxygen vacancies states and the valence band [20].

The PL intensity from the nanocomposite CdS+ZnO films is about an order of magnitude higher than that from the similarly deposited pure (uncapped) nanocrystalline CdS or ZnO films with similar mean particle size. We ascribe the PL enhancement to the presence of ZnO nanoparticles in the nanocomposite that effectively quench non-radiative decay processes. Also, unlike the pure nano-CdS films which need prolonged post deposition surface treatment with a $H_2+N_2$ mixture to prevent chemical degradation [17], the as-deposited CdS+ZnO films are inherently inert to atmospheric degradation, and do not require such treatment. Note also that in pure CdS, there is a growth in particle size in the sputtered films with increasing film thickness. However, the presence of two components in the nanocomposite film effectively retards the particle growth for either phase.

We also carried out time-resolved fluorescence measurements using the second harmonic from a Ti-sapphire laser (440nm, 1ps) as excitation source. The majority species in the CdS+ZnO nanocomposite was found to have a lifetime <50ps (as obtained from an exponential fit), while both nano-CdS (166ps) and nano-ZnO (329ps) had substantially longer lifetimes. The short decay time in the nanocomposite appears to be a direct consequence of the enhanced oscillator strength in this system. The enhanced PL and fast relaxation are both desirable for device applications.

## 6. Coherence

It is clear from Fig. 2 that the PL emission from the nanocomposite sample shows sharp intensity modulations. Since the fringe spacing is found to be inversely proportional to the film thickness, such oscillations can be ascribed to multiple-beam, thin-film interference. Intensity modulations are also seen (Fig. 2c) in the thicker nano-CdS film ($t = 1.3\mu m$) but not in the thinner one (Fig. 2d, $t = 0.2\mu m$), because the fringe spacing is comparable to the PL spectral width in the latter case. However, a nano-ZnO film of sufficient thickness ($t =$ 0.9μm) *does not* exhibit any intensity modulation (Fig. 2e). We point out that apart from external parameters such as the quality and reflectivity of the film-substrate interface, the fringe contrast in multiple beam interference is governed by the degree of spatial and temporal coherence of the light source.

Though such fringes should technically be observed whenever the film thickness is less than the temporal coherence length, the fringe contrast would actually depend on the extent of spatial coherence. We estimated the *degree* of spatial coherence in the light emitted from nanoparticle ensembles using a version of Young's double slit experiment [5], which is a time-honored test for the existence of spatial coherence. The PL was excited by a Ti:Sapphire laser (400nm, 80fs) focused to a spot size of ≈2mm on the sample. A filter was used to block the scattered laser radiation from the detector. Fig. 3 shows the two-beam interference pattern produced by the PL emitted by the CdS-ZnO nanocomposite.

The degree of spatial coherence, ($|j_{12}|$), is a complex quantity related to the fringe contrast in the double slit experiment. A value of $|j_{12}| = 1$ corresponds to perfect coherence, that is 100% contrast in the fringe pattern. We calculated $|j_{12}|$ by fitting the data (Fig. 3) to an equation [21] of the form: $I(Q) = I_1(Q) + I_2(Q) + 2\sqrt{I_1(Q)I_2(Q)}|j_{12}|f(Q,r)$, where $I(Q)$ is the intensity at the point Q, $I_{1,2}$ is the intensity passing through the first (second) slit and $f(Q,r)$ is a function of the difference in the separation of the two slits from $Q$. The fitted curve shown in Fig. 3 gave $|j_{12}| = 0.2$ for the CdS-ZnO nanocomposite. For an extended source emitting spontaneously (as against stimulated emission), the observed degree of coherence is unexpectedly high. The coherence can be further enhanced by reducing the width of the PL emission profile, which may presumably be achieved by producing a sample with a narrower particle size distribution.

Note that the fringe contrast and $|j_{12}|$ are expected to deteriorate with an increase in the angular dimension of the source, and the spatial coherence is finally lost above a certain

source size. The characteristic length scale over which it is lost is its spatial coherence length. The relation between the fringe contrast and the size of the illuminated spot on the sample was studied by recording PL spectra for two different spot sizes (2mm×1mm and 20μm×10μm). For the CdS-ZnO nanocomposite (Fig. 4, top), the spectra corresponding to both spot sizes show modulation but the contrast is expectedly better for the smaller spot size. Nano-ZnO, on the other hand shows no modulation at 2mm×1mm and a very weak modulation at 20μm×10μm (Fig. 4, bottom). For the present experimental geometry, the spatial coherence length of the nanocomposite ensemble can be calculated [21] to be ≈10μm, which is much larger than the corresponding quantity in nano-ZnO (≈350nm). The PL emission from nano-CdS is too weak to allow an accurate measurement of the coherence length, but it was estimated to have a value in between those of the nanocomposite and nano-ZnO.

## 7. Conclusions

The electronic properties of a nanocomposite of two different components (in which there is an appreciable overlap of wavefunctions between particles) cannot be expressed simply as a linear superposition of the properties of the individual components. This opens up a region of phase space with crystallite size and component fraction as coordinates. The CdS-ZnO nanocomposite exhibits much higher PL emission (without any additional surface passivation being required), larger spatial coherence length and shorter decay times than individual CdS or ZnO nanocrystals of similar dimensions. The nanocomposite structure can also be formed more simply and efficiently than the core-shell structure.

It is a pleasure to thank G. Krishnamoorthy for the fluorescence lifetime measurements, A. S. Vengurlekar and K. L. Narsimhan for stimulating discussions, and S. Bhattacharya for his advice and a critical reading of the manuscript.

# References

[1] L. Brus, *Appl. Phys. A - Materials Science & Processing* **53** (1991) 465.

[2] V. Ranjan, V. A. Singh and G. C. John, *Phys. Rev. B* **58** (1998) 1158.

[3] H. Cao, Y. G. Zhao, H. C. Ong, S. T. Ho, J. Y. Dai, J. Y. Wu, and R. P. H. Chang, *Appl. Phys. Lett.* **73** (1998) 3656; H. Cao, Y. G. Zhao, S. T. Ho, E. W. Seelig, Q. H. Wang, and R. P. H. Chang, *Phys. Rev. Lett.* **82** (1999) 2278.

[4] M. Gurioli, F. Bogani, S. Ceccherini, and M. Colocci, *Phys. Rev. Lett.* **78** (1997) 3205.

[5] T. Young,, *Phil. Trans. Roy. Soc., London* xcii **12** (1802) 387.

[6] U. Woggon, *Optical Properties of Semiconductor Quantum Dots* (Springer, Berlin) 1997.

[7] S. V. Gaponenko, *Optical Properties of Semiconductor Nanocrystals* (Cambridge University Press, Cambridge) 1998.

[8] M. V. Artemyev, A. I. Bibik, L. I. Gurinovich, S. V. Gaponenko, and U. Woggon, *Phys. Rev. B* **60** (1999) 1504.

[9] V. A. Shchukin, D. Bimberg, T. P. Munt, and D. E. Jesson, *Phys. Rev. Lett.* **90** (2003) 076102.

[10] A.P Alivisatos, *J. Phys. Chem.* **100** (1996) 13226.

[11] B. G. Potter Jr. and J. H. Simmons, *Phys. Rev. B* **37** (1988) 10838.

[12] Y. Wang and N. Herron, *J. Phys. Chem.* **91** (1987) 257.

[13] H. Mathieu, T. Richard, J. Allègre, P. Lefebvre, G. Arnaud, W. Granier, L. Boudes, J. L. Marc, A. Pradel and M. Ribes, *J. Appl. Phys.* **77** (1995) 287.

[14] N. Chestnoy, T. D. Harris, R. Hull and L. E. Brus, *J. Chem. Phys.* **90** (1986) 3393.

[15] R. Banerjee, R. Jayakrishnan and P. Ayyub, *J. Phys. - Cond. Mat.* **12** (2000) 10647.

[16] P. Ayyub, R. Chandra, P. Taneja, A. K. Sharma and R. Pinto, *Appl. Phys. A* **73**, 67 (2001).

[17] P. Taneja, P. Vasa, and P. Ayyub, *Mater. Lett.* **54**, 343 (2002).

[18] S. V. Nair, L. M. Ramaniah, and K. C. Rustagi, *Phys. Rev. B* **45**, 5969 (1992).

[19] J. R. Lakowicz, I. Gryczynski, Z. Gryczynski, and C. J. Murphy, *J. Phys. Chem. B* **103**, 7613 (1999).

[20] M. Haase, H. Weller, and A. Henglein, *J. Phys. Chem.* **92**, 482 (1988).

[21] M. Born and E. Wolf, *Principles of Optics*, **7**th edition (Cambridge University Press, Cambridge, 1999), Chap. 10.

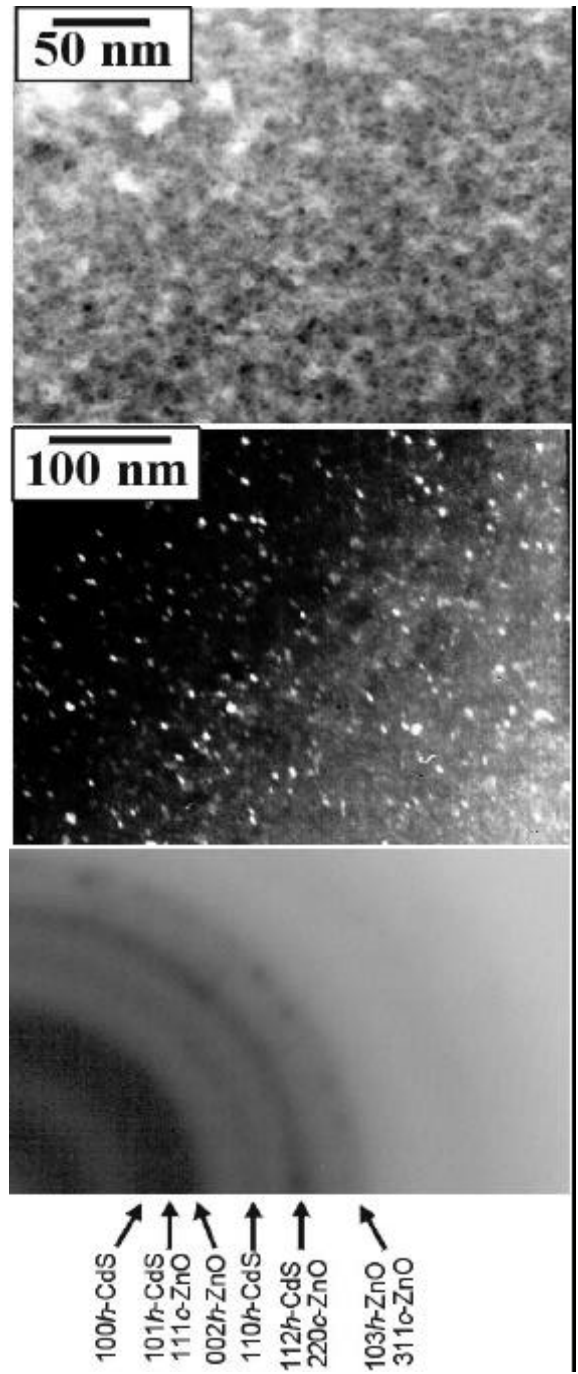

**Figure 1.** (Top) Bright-field TEM micrograph of nanocomposite CdS+ZnO thin film sputter-deposited on Si. The micrographs show two distinct, nanodispersed phases with different contrasts, each phase consisting of crystallites of average size 2-3nm. (Center) Dark-field TEM micrograph from the same area, highlighting only a fraction of the particles that contribute to a segment of a single diffraction ring. (Bottom) Selected area electron diffraction pattern from the same region, indexed on the basis of the hexagonal (*h*) and cubic (*c*) polymorphs of CdS and ZnO.

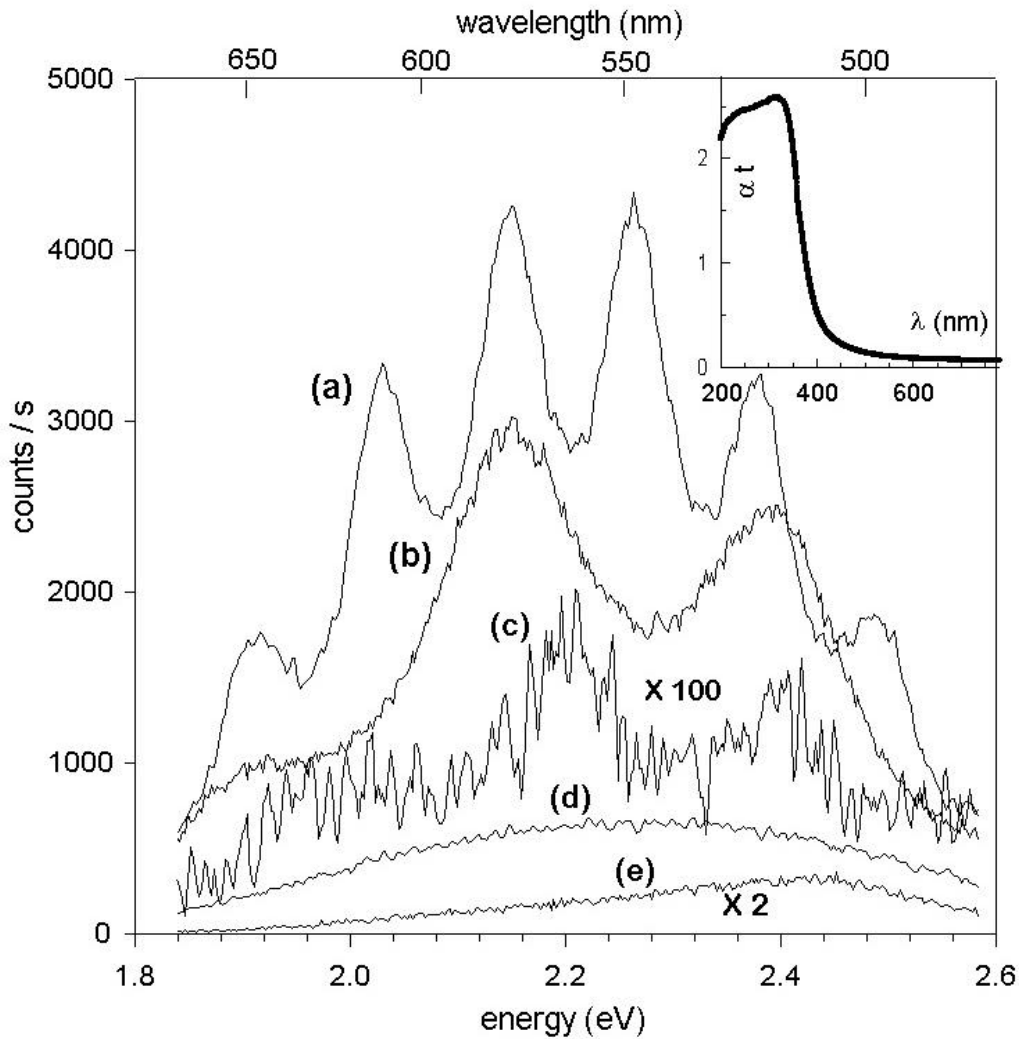

**Figure 2.** Photoluminescence emission spectra recorded with an excitation wavelength of 457.9nm (Ar-laser) from (a) CdS+ZnO nanocomposite film ($t$ = 1.8μm, $d_{av}$ = 2.8nm), (b) CdS+ZnO nanocomposite film ($t$ = 0.9μm, $d_{av}$ = 2.8nm), (c) nanocrystalline CdS film ($t$ = 1.3μm, $d_{av}$ = 5.2nm), (d) nanocrystalline CdS film ($t$ = 0.2μm, $d_{av}$ = 2.4nm), and (e) nanocrystalline ZnO film ($t$ = 0.9μm, $d_{av}$ = 3nm). Here $t$ = film thickness and $d_{av}$ = mean particle size obtained from the quantum shift of the optical band gap. Curve (c) has been multiplied by 100 and curve (e) has been multiplied by 2. Inset shows the optical absorption spectrum from a typical CdS+ZnO nanocomposite film deposited on quartz.

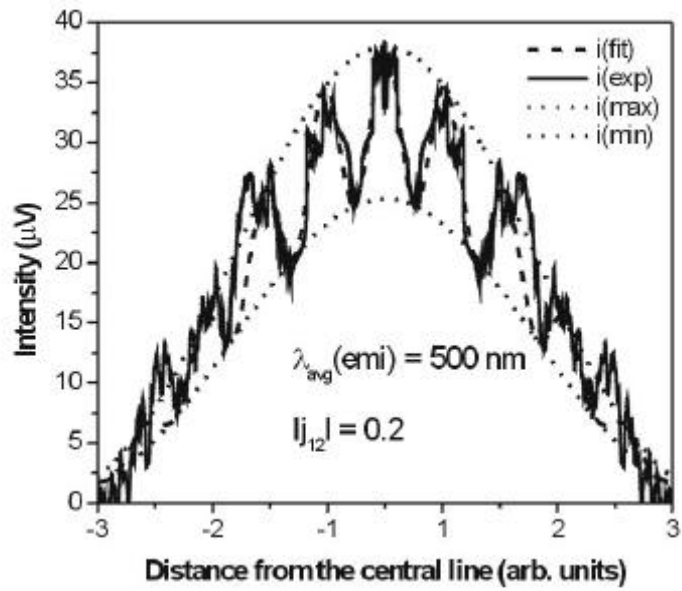

**Figure 3.** Two beam interference pattern generated from the photoluminescence emitted by a CdS-ZnO nanocomposite film (average emission wavelength = 500nm) in a Young's double slit geometry.

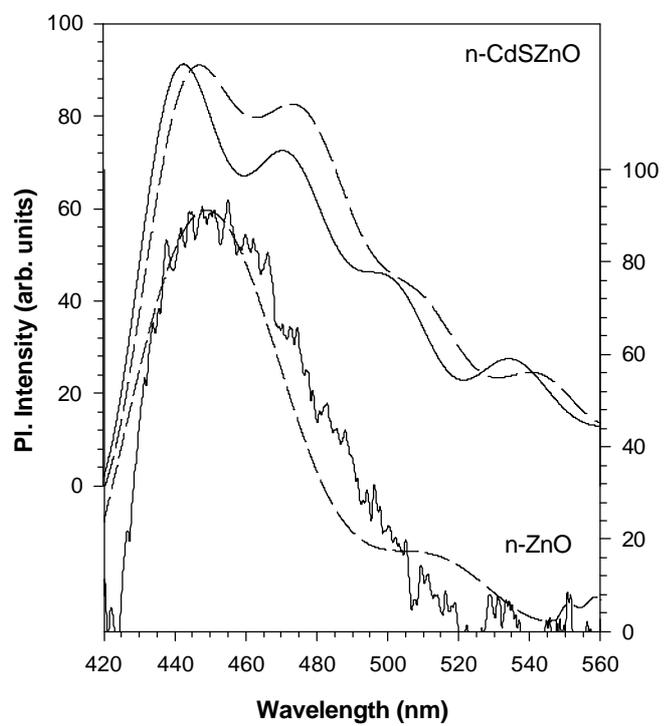

**Figure 4.** Photoluminescence spectra from thin films of nanocomposite CdS-ZnO (top) and nanocrystalline ZnO (bottom) for excitation spot sizes of 20μm×10μm (solid line) and 2mm×1mm (dashed line).